\documentstyle[multicol,aps]{revtex}
\def\baselinestretch{1.1}

\newcommand{\be}{\begin{equation}}
\newcommand{\ee}{\end{equation}}
\newcommand{\half}{\frac{1}{2}}

\renewcommand{\narrowtext}{\begin{multicols}{2} \global\columnwidth20.5pc}
\renewcommand{\widetext}{\end{multicols} \global\columnwidth42.5pc}

\begin{document}

\title{\Large\bf 
Remarks on the black hole entropy and 
Hawking spectrum in Loop Quantum Gravity
}
\author{ \large\bf A. Alekseev$^{1,3}$, 
A. P. Polychronakos$^{2}$, M. Smedb\"ack$^{3}$ \\[2mm] }

\address{\noindent $^{1}$  
University of Geneva, Section of Mathematics, 2-4 rue du Li\`evre,
1211 Gen\`eve 24, Switzerland\\
$^{2}$ Physics Department, City College of the CUNY,
138th Street and Convent Avenue, New York, NY 10031, USA\\
$^{3}$ Department of Theoretical Physics, Uppsala University,
Box 803, SE-75108, Uppsala, Sweden\\
e-mails: alekseev@math.unige.ch, alexios@sci.ccny.cuny.edu,
mikael@teorfys.uu.se \\
 }

\date{May 2004} 

\maketitle

\begin{abstract}
\noindent

In this note we reply to the criticism by Corichi concerning 
our proposal for an equidistant area spectrum in loop quantum gravity.
We further comment on the emission properties of black holes and 
on the  statistics of links.

\vspace{1mm}
\noindent
{PACS-98: 04.70.Dy, 11.15.-q \hfill CCNY-HEP-04/3 ~~ UUITP-13/04}
\end{abstract}

\narrowtext


This note consists of three comments on different aspects
of the loop quantum gravity approach to the black hole physics.

{\em 1. Area spectrum.}
The standard result in loop quantum gravity is that the area quantum
of a spin-$j$ link is proportional to the square root of its Casimir
\cite{spectrum}:
\be\label{trad}
A_j = \gamma \sqrt{j(j+1)}
\label{irra}
\ee
In the previous paper \cite{APS}, we have argued that a different
regularization of
the area operator produces an additive quantum
correction of $1/4$ to the Casimir, leading to the equidistant
area spectrum:
\be\label{aps_spectrum}
A_j = \gamma (j+\half)
\label{equi}
\ee
This spectrum conforms with early work of Bekenstein and Mukhanov
 \cite{BekMuk},
who argued for an equidistant area spectrum for black
hole horizons, and naturally leads to a composition of Schwarzschild
black holes in terms of 3-state objects (spin-1 links), in agreement
with recent heuristic observations based on quasi-normal black hole
modes \cite{Pol}. 

Recently, Corichi \cite{Cori} has criticized the spectrum 
(\ref{aps_spectrum}) as inconsistent with
the principles of loop quantum gravity. His argument amounts to the
statement that a link carrying the singlet (spin-0) representation
should be considered as part of the vacuum and be equivalent to no
link at all; therefore, its quantum of area should be zero.
For completeness, let us  mention that 
there is yet another regularization of the area operator \cite{AFS} with an 
equidistant spectrum  $A_j = \gamma j$ that satisfies Corichi's conditions.

Our approach to this issue is as follows. Loop quantum gravity 
is a reformulation of the original Einstein gravity. As such,
it is non-renormalizable. The definition of such a theory
depends on the regularization scheme and there is no
scheme that produces cutoff-independent results in the infrared
region.  In this situation, different regularizations have an equal claim, 
and will in general produce different physics.

From this perspective, various alternative
formulations of loop quantum gravity, such as random lattice or spin foam
theory, are independent starting points that are simply motivated by the
original theory. One possible point of view is that
spin links are essentially quantum
mechanical variables which can be created or annihilated. Corichi insists
that the creation of a spin-0 link should be equivalent to no link at
all. Instead, we propose that the presence of a spin-0 link is a
physically distinct event, altering the connectivity of the lattice.
We assign to such a spin nontrivial quantum numbers, like the area.

An ultimate test for a disretized gravity model 
is whether or not it reproduces the correct general 
relativistic infrared physics. This is as yet unclear
for all currently available modifications of loop
quantum gravity. Hence, at the
present level of our understanding of quantum gravity, all the above
formulations should be explored and be placed on their proper
semi-heuristic footing.

{\em 2. Link statistics.}
Next, we comment on the statistics of the links in the discretized
gravity models. As pointed out in
\cite{APS}, the standard picture of macroscopic black holes as condensates
of links with a particular value of spin (half or one) holds if the links
are considered only partially distinguishable. Fully indistinguishable
links fail to reproduce the area-entropy relation (one gets
$S \sim A^t$ with $t<1$). Fully distinguishable
links, however, give an entropy proportional to the area, although with
no single-spin condensate.

In a recent paper, Khriplovich \cite{Khr} states that fully distinguishable
links would overestimate the entropy as $S \sim A \ln A$. We point out that
this is an erroneous result, arising from a misunderstanding of the notion of
distinguishability. Links could become distinguishable due to their different
placement and connectivity in the spacetime lattice. This means that states
with the (distinct) spin quantum numbers of two links exchanged are
distinct. Khriplovich effectively considers also exchanging their position,
which was the reason that made them distinguishable in the first place,
thus erroneously overcounting the states. For example, a state with two
distinct spin-half links, both in a spin-up state, is unique, while
Khriplovich would count it as two states, exchanging a fictitious label
between the two links. The correct counting was done in \cite{APS} and
yields $S \sim A$.

{\em 3. Hawking spectrum.}
We also take this opportunity to comment on the oft-stated claim that
the equidistant area spectrum (\ref{equi}) predicts emission in discrete
lines, while the standard, irrational, area spectrum (\ref{irra}) for
links would reproduce a continuous Hawking radiation spectrum.

In fact, this is far from clear. 
In the standard statistics of links, a macroscopic black hole 
is dominated
 by a condensate of
links of the same spin (spin-half for the standard spectrum, spin-one
for our spectrum), with only a small number of links of different
spins. Thus, the black hole evaporation process 
will be dominated
by events where the number of links in the condensate decreases (links
disappear). All such links carry the same area quantum and decrease
the energy by the same discrete amount.

Krasnov \cite{Kra} has provided a detailed analysis of the 
black hole evaporation in loop quantum gravity. He suggests
the thermal spectrum, taking into account the thermodynamic dominance of one
kind of links (spin-half), but arguing that the process of annihilation
of a spin-half link (transition to spin zero, for the standard spectrum)
is forbidden due to gauge invariance (singlet condition).
It is then argued that the remaining processes of de-excitement
of higher-spin links would produce the continuous radiation spectrum.

We point out that, although decay of a single spin-half link is
forbidden, the simultaneous decay of {\it two} such links
is allowed and will be the process that thermodynamically overwhelms
black hole evaporation. At any rate, after the few existing higher-spin
links have de-excited as Krasnov proposes, there is no way for the
black hole to further reduce its area except by ``shedding'' links of
the condensate, in the process described above. There is no escape
from the fact that the evaporation process cannot rely on single-decay
events and will largely involve simultaneous transitions of two 
spin-half links. 

The exact emission spectrum will depend on the
preferred channels of such transitions. For instance, we could consider
the two-stage decay (parentheses denote links with the corresponding spin)
\be\label{decay}
(\half) + (\half) \to (1) \to 0
\ee
which could in principle produce
intermediate lines. If, however, the second, single-link
decay process is much faster than the first, and faster than the typical
inverse energy gap between states, by the standard time-energy
uncertainty principle it is quantum mechanically unobservable and the
whole process is equivalent to one-stage decay of two spin-half links
with a single spectral line.

Let us also point out that a discrete line spectrum would predict a
black hole evaporation rate drastically slower than  the Hawking one.
Indeed, discrete lines correspond to highly stable energy states and 
therefore low emission intensity. In more detail, a discrete area
spectrum with steps of order 1 in Planck units, equidistant or not, would
imply discrete energy levels with steps of order $M^{-1}$, where
$M \sim \sqrt{A}$ is the black hole mass. Assuming a decay rate of the
Hawking order of magnitude, ${\dot M} \sim M^{-2}$, would imply a lifetime
for each of these levels of order $M$ and a corresponding spread of order
$M^{-1}$, completely washing out their discrete nature.
In conclusion, the discrete spectrum and the Hawking radiation rate 
seem to be incombatible with each other.

The discussion above does not take into account the
influence of matter. The latter can eventually smooth
and thermalize the emission spectrum, but at the moment we have no
control over such effects.

{\bf Acknowledgements}. A.A. acknowledges the support of
the Swiss National Science Foundation. The research of A.P. was
supported in part by NSF grant NSF-0353301.

\def\baselinestretch{1.0}

\widetext

\end{document}